\documentclass[twocolumn,secnumarabic,amssymb, nobibnotes,aps,prl,superscriptaddress]{revtex4-2}
\usepackage{graphicx}
\usepackage{textcomp}
\usepackage{amsmath}
\usepackage{commath}
\usepackage{color}
\usepackage{appendix}

\newcommand{\sg}[1]{}
\renewcommand{\sg}[1]{{\color{blue}{#1}}} 

\begin{document}

\title{Stress focusing and damage protection in topological Maxwell metamaterials}
\author{Caleb Widstrand}
\affiliation{Department of Civil, Environmental, and Geo- Engineering, University of Minnesota, Minneapolis, MN 55455, USA}
\author{Chen Hu}
\affiliation{Department of Civil, Environmental, and Geo- Engineering, University of Minnesota, Minneapolis, MN 55455, USA}
\author{Xiaoming Mao}
\affiliation{Department of Physics, University of Michigan, Ann Arbor, MI 48109, USA}
\author{Joseph Labuz}
\affiliation{Department of Civil, Environmental, and Geo- Engineering, University of Minnesota, Minneapolis, MN 55455, USA}
\author{Stefano Gonella}
\email{sgonella@umn.edu}
\affiliation{Department of Civil, Environmental, and Geo- Engineering, University of Minnesota, Minneapolis, MN 55455, USA}

\begin{abstract}
    \noindent Advances in the field of topological mechanics have highlighted a number of special mechanical properties of Maxwell lattices, including the ability to focus zero-energy floppy modes and states of self-stress (SSS) at their edges and interfaces.
    Due to their topological character, these phenomena are protected against perturbations in the lattice geometry and material properties, which makes them robust against the emergence of structural non-idealities, defects, and damage. 
    Recent computational work has shown that the ability of Maxwell lattices to focus stress along prescribed SSS domain walls can be harnessed for the purpose of protecting other regions in the bulk of the lattice from detrimental stress concentration and, potentially, inhibiting the onset of fracture mechanisms at stress hot spots such as holes and cracks. 
    This property provides a powerful, geometry-based tool for the design of lattice configurations that are robust against damage and fracture. 
    In this work, we provide a comprehensive experiment-driven exploration of this idea in the context of realistic structural lattices characterized by non-ideal, finite-thickness hinges.
    Our experiments document the onset of pronounced domain wall stress focusing, indicating a remarkable robustness of the polarization even in the presence of the dilutive effects of the structural hinges.
    We also demonstrate that the polarization protects the lattice against potential failure from defected hinges and cracks in the bulk. 
    Finally, we illustrate numerically the superiority of SSS domain walls compared to other trivial forms of reinforcements.
		\vspace{0.4cm}
\end{abstract}	\maketitle

\section{Introduction}

Metamaterials are engineered solids with specially designed qualities, defined by their microstructure, that are not typically found in nature~\cite{CuiMeta_2010}.
In particular, \textit{mechanical metamaterials} are designed to display unconventional mechanical properties and exhibit unique responses to external loading~\cite{Bertoldi_2017,Surjadi_Meta_2019}. 
Several recent developments in the field of metamaterials have arisen from the injection of the notion of topological mechanics in the design of architected media.
A class of systems of great interest in topological mechanics is based on the so-called topological Maxwell lattice. 
Maxwell lattices have the same number of constraints and degrees of freedom in the bulk~\cite{Maxwell_1864}, thus being on the verge of mechanical instability.
In two dimensions, this class of lattice includes square and kagome lattices~\cite{Phani2006,Souslov2009,MaoStenullLubensky_Filamentous_2013}.
In the presence of open boundary conditions (i.e., in a finite domain), topological Maxwell lattices have the ability to \textit{localize their response} along their boundaries, making them a mechanical analogue to topological insulators in quantum electronic systems~\cite{KL_2013}.
Specifically, they can localize zero-frequency floppy modes along what is referred to as a ``floppy edge," while leaving the opposite edge rigid, ultimately displaying an asymmetric mechanical response~\cite{KL_2013,Rocklin_2017_NJP,Mao_Lubensky_2018,Baardink_PNAS_2018,ZhouZhangMao_Fibers_2018,ZhouZhangMao_Quasicrystals_2019}. 
This topological behavior can be quantified by a polarization vector that defines the direction along which floppy modes tend to exponentially localize.
The bulk-edge correspondence implies that the polarization is not an intrinsic property of the edges; rather it is a property of the bulk that manifests at the edges.
Therefore, it is guaranteed to persist as long as the topological properties of the bulk are not altered; this immunity is referred to as topological protection~\cite{Paulose_2015,Zhang_Fracture_2018}.
The ability to localize deformation at the edges results in an ability to absorb localized loads without transferring significant stresses into the bulk, a property that can be exploited for static cloaking of sharp objects~\cite{Zunker_2021} and impact protection.

\begin{figure*}[t]
	\centering
	\includegraphics[width=1\textwidth]{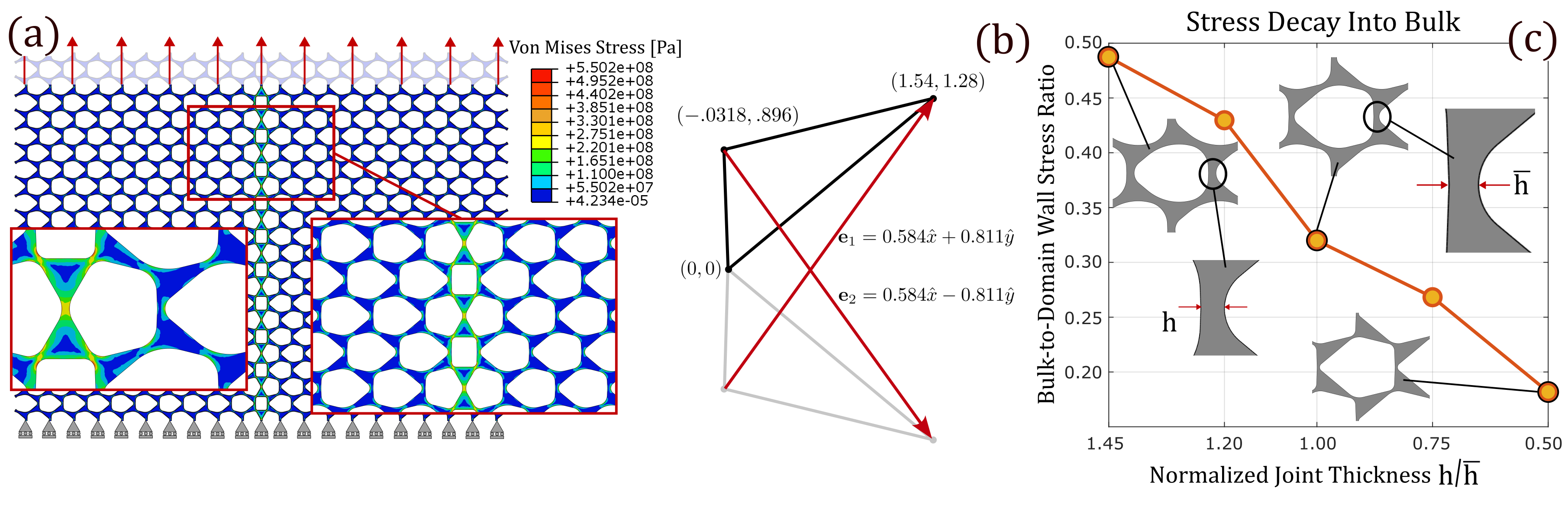}
	\caption{(a) Lattice with SSS domain wall under tensile boundary conditions; a row of unit cells at the top is uniformly translated axially, and roller boundary conditions restricting axial displacement are applied at the bottom. Colormap shows von Mises stress. (b) Idealized unit cell geometry with coordinates of the upper triangle marked in centimeters. Primitive lattice vectors are also marked. (c) Evolution of bulk-to-domain wall stress ratio as a function of normalized hinge thickness. The trend shows a drop in focusing ability going from thin to thick hinges; nevertheless, even for thick hinges, a considerable level of focusing is preserved.} \label{IdealBC}  
\end{figure*}

Recently, relevant work has been devoted to study the resilience of mechanical and topological properties in the transition from ideal configurations featuring perfect hinges to \textit{structural} lattices, as they would be obtained via machining, cutting, or 3D-printing.
Specifically, it has been shown that the topological polarization is preserved, albeit diluted in strength~\cite{Pishvar_Harne_2020,Chen_Huang_2022}, as a result of the additional interactions between unit cells that over-constrain the lattices.
These kinematics cause the boundary modes predicted for ideal Maxwell lattices to migrate to finite frequencies in the form of floppy edge phonons~\cite{JihongMa_2018,Stenull_2019,Charara_Bilayer_2021}.
More directly relevant to this work are a number of experimental efforts that have characterized the static mechanical response of structural Maxwell lattices. 
Paulose et al. investigated layered kagome lattices and experimentally demonstrated the onset of selective buckling behavior controlled by the availability of domain walls~\cite{Paulose_2015}.
Work by Chapuis et al. experimentally characterized the material properties of topological Maxwell beam networks capable of localizing stress along non-linear interfaces~\cite{Chapuis_Shea_2022}. 
Chen et al. used microtwist modeling to characterize hinged kagome lattices~\cite{Chen_Huang_2022}.
Due to the reconfigurability that they easily undergo through simple relative deformations or relative rotations of the triangles, kagome lattices have also been studied for their potential as transformable metamaterials. 
In this vein, Rocklin et al. were able to track the evolution of their topological behavior as a function of changes in the twist angle through phase maps that revealed the emergence of topological phase transitions~\cite{Rocklin_2017}.

In parallel to floppy modes, Maxwell lattices can also support \textit{states of self-stress} (SSS), i.e., self-equilibrated states of stress in the bonds that do not result in net forces at the lattice sites~\cite{KL_2013,Paulose_2015,Chapuis_Shea_2022}.
When we consider lattices with periodic boundary conditions (infinite lattices), SSS can develop in the bulk, and are associated with the existence of directions of aligned bonds, or fibers, as observed for instance along three major directions in the regular kagome lattice~\cite{Mao_Lubensky_2018}.
For the case of finite lattices, a condition of special interest is the localization of SSS at interfaces (or domain walls) between lattice domains.
In this vein, Kane and Lubensky showed that zero modes localize at the interfaces between differently polarized lattices~\cite{KL_2013}.
Specifically, interfaces obtained by stitching a floppy edge to another floppy edge, or to a trivial non-polarized region, result in a soft domain wall.
Conversely, interfaces obtained by stitching a rigid edge to another rigid edge, or to a non-polarized region, localize SSS and focus stress. 
Paulose et al. were the first to demonstrate experimentally the potential of SSS domain walls as effective stress guides~\cite{Paulose_2015}.

Zhang and Mao further investigated SSS of topological kagome lattices for their stress focusing and fracture protection potential~\cite{Zhang_Fracture_2018}.
They noted that such stress localization has an implication of major engineering significance: when a conventional material with a crack, defect, or sharp notch is mechanically loaded, the stress inevitably focuses at their tips, making them mechanical hot spots that can potentially lead to catastrophic failure. 
In contrast, when a topological Maxwell lattice with an SSS domain wall is loaded, the stress tends to focus predominantly on the domain wall even in the presence of defects, holes, or cracks in the bulk. 
As a result, the severe stress concentration effect classically observed at a crack tip can be avoided or significantly reduced. 

The main objective of this study is to assess experimentally the robustness of stress focusing at SSS domain walls in structural lattices featuring hinges with finite thickness.
Along with previous experimental studies, this investigation closes the loop on the question of whether the topological phenomena of Maxwell lattices survive in the presence of the structural conditions naturally encountered in working with realistic hinged lattices and, if so, to what degree of dilution.
The other part of our study is devoted to showing how the topological attributes of the lattice set them apart from other mechanical systems in which similar protection may be sought via fiber reinforcement strategies~\cite{Ozgur_FiberBook_2017,Giorgio_fiber_2020}.

\begin{figure*}[t]
	\centering
	\includegraphics[width=1\textwidth]{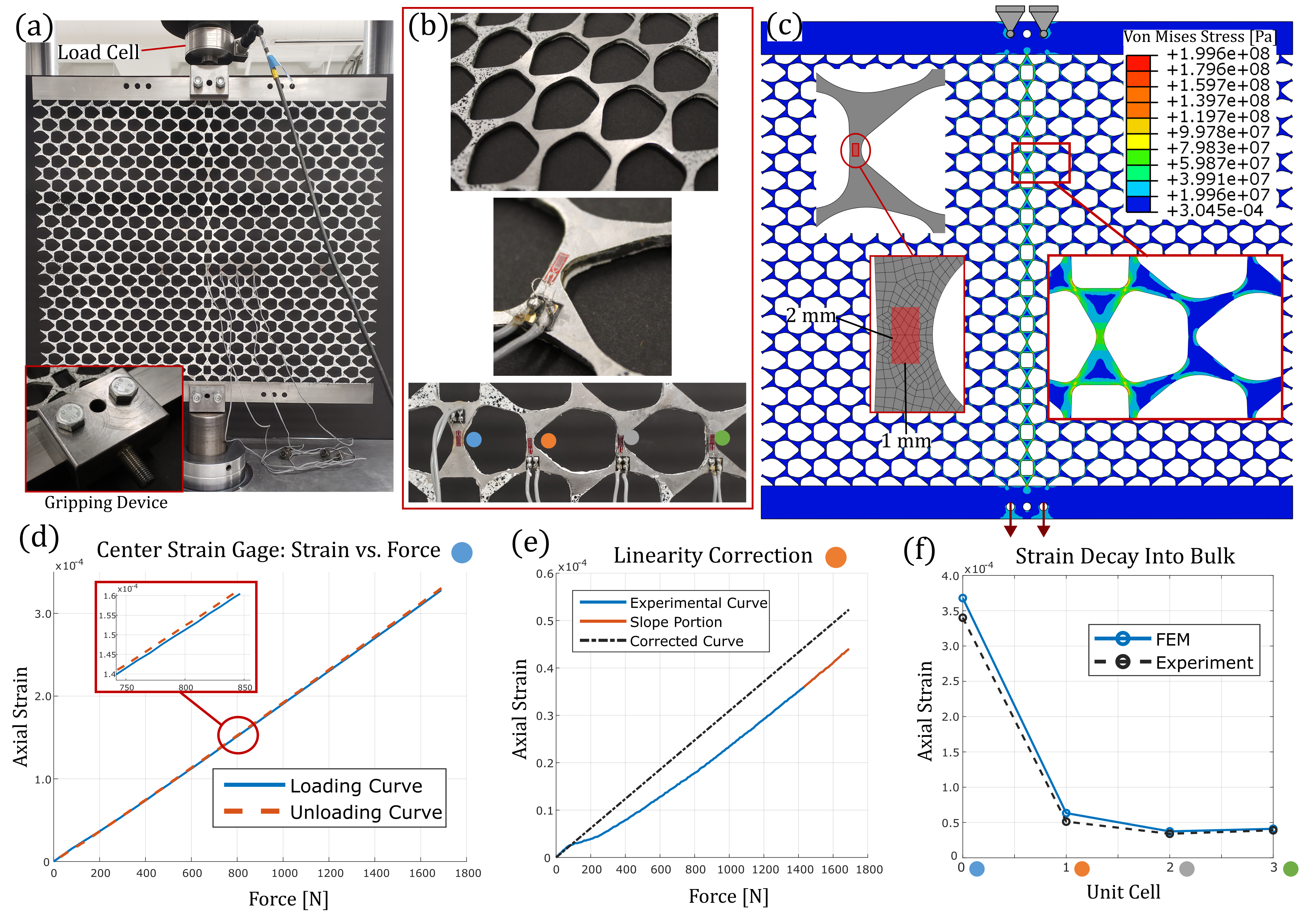}
	\caption{(a) SSS domain wall specimen set in MTS load frame, with details of load cell and gripping device with bolted connections. (b) Top-to-bottom: Close-up of unit cells. Close-up of a single strain gage. Close-up of four strain gages that are used to monitor decay of axial strain into the bulk with colored markers. (c) FEM model of lattice matching geometry and pinned boundary conditions of specimen and axial load applied to the holes to mimic action of grippers. The stress map shows strong localization at the SSS domain wall. Close-up of hinge where the strain is measured by strain gage in the physical specimen, with inset showing the nodes falling in the area of the strain gage, whose values are averaged to extract the measure to compare against the gage measurement. (d) Plot of axial strain in the center gage vs. force showing linearity and matching loading and unloading paths. (e) Experimental strain-force curves displaying nonlinearities in the early loading stages to be attributed to adjustments in the frame-specimen setup until sufficiently large values of force are reached and calling for a slope-correction procedure. (f) Axial strain in the hinges of unit cells progressively moving away from the the domain wall: experimental measurements from the strain gages and numerical estimates from the FEM model are in excellent qualitative agreement, capturing a sharp decay rate into the bulk while also displaying remarkable quantitative agreement.} \label{CenteredExpModel}  
\end{figure*} 

\section{Stress focusing at SSS domain walls}

To first assess the ability of Maxwell metamaterials to focus stress, we construct a two-dimensional continuum elasticity model featuring two oppositely polarized regions of unit cells, as shown in Fig.~\ref{IdealBC}(a). 
The unit cell for the selected configuration, in its idealized form, is pictured in Fig.~\ref{IdealBC}(b) with the coordinates for the upper triangle marked in centimeters; the primitive lattice vectors are also marked.
The polarization is determined using the criterion provided in \cite{KL_2013}, which allows us to calculate the direction of the polarization vector using only the primitive kagome lattice vectors ($\mathbf{a}_p$) and the deflections of the sites of the unit cell relative to the sites of a regular kagome unit cell ($x_p$):
\begin{equation}
    \mathbf{R}_{T} = \sum_{p=1}^3 (\mathbf{a}_p sgn~x_p)/2
\end{equation}
Here, the configuration is polarized in the horizontal direction, making it amenable to mirroring about the vertical axis and stitching.
The stitching operation is enabled by the introduction of a one cell-wide layer of modified cells to properly join the two subdomains, following a protocol successfully used in previous works~\cite{,Paulose_2015,Zhang_Fracture_2018,Chapuis_Shea_2022}.
A finite element model (FEM) of the lattice, 20 unit cells wide by 12 unit cells tall, is assembled in ABAQUS, using a mesh of two-dimensional, four-noded plane-stress elements with appropriate mesh refinement at the hinges.

\begin{figure*}[t]
	\centering
	\includegraphics[width=1\textwidth]{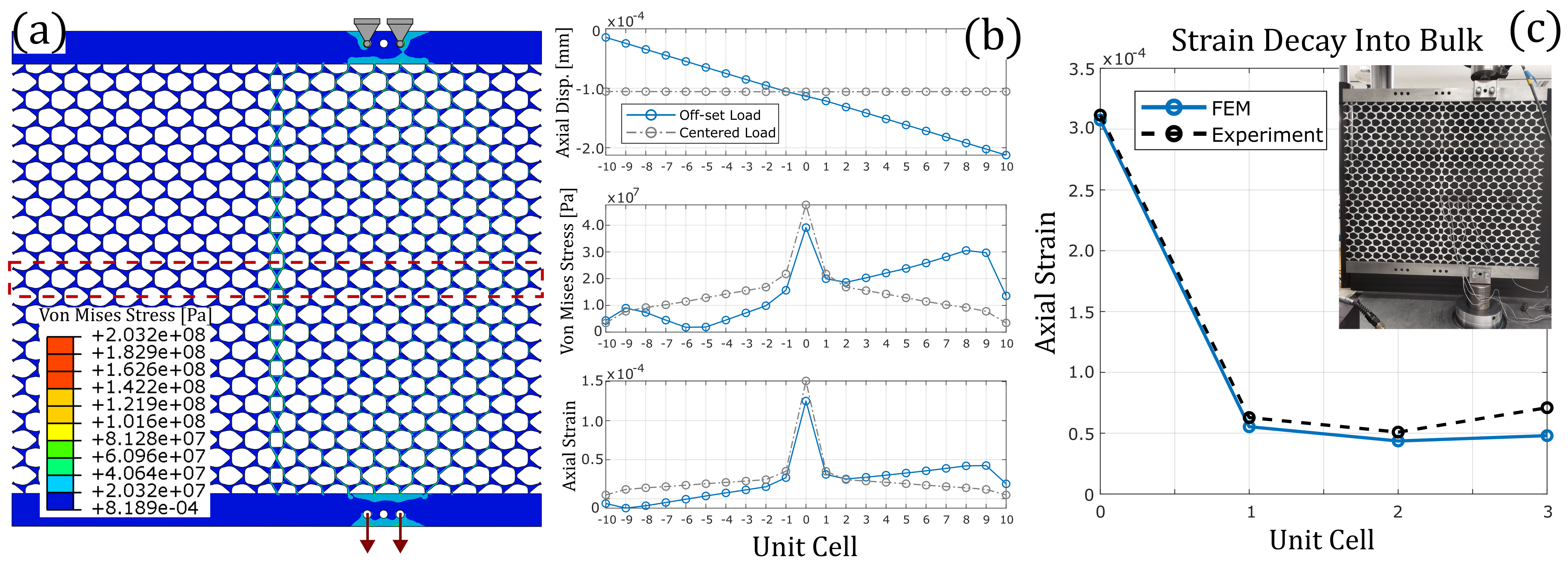}
	\caption{(a) FEM model of lattice with offset loading. Dashed box marks sampling region for stress. (b) Displacement, von Mises stress, and axial strain (cell averages) plotted against cell index for cells inside highlighted row, for both the offset and centered loading cases. Strains and stresses clearly show peaks at the SSS domain wall despite the offset. (c) Decay of axial strain from the domain wall into the bulk of the lattice, displaying excellent agreement between experiments and numerics. Inset photo shows specimen in the load frame in the offset configuration, for reference. } \label{OffsetExperiment}
\end{figure*}

The model, shown in Fig.~\ref{IdealBC}(a), implements boundary conditions mimicking as closely as possible those used on ideal Maxwell lattices in~\cite{Zhang_Fracture_2018}, where an axial displacement is applied uniformly to the top row of unit cells and a roller boundary condition restricting only axial displacement is applied along the bottom row. 
The hinge thickness is $\Bar{h} = 2~mm$ in Fig.~\ref{IdealBC}(a).
Linear static analysis reveals a significant amount of stress focusing occurring at the SSS domain wall of the lattice, with most of the decay into the bulk occurring within one cell from the interface. 
In order to assess this degree of dilution of the ability to focus stress introduced by the finite thickness of the structural hinges, we repeat the analysis for several hinge thicknesses h, as shown in Fig.~\ref{IdealBC}(c).
For each, we compute a bulk-to-domain wall stress ratio by taking the von Mises stress averaged over the unit cells in the bulk (starting one cell away from the domain wall) and dividing it by a corresponding quantity averaged along the domain wall.
We plot this stress ratio against a normalized hinge thickness in Fig.~\ref{IdealBC}(c). 
As expected, lower thicknesses approaching the ideal limit cause a greater concentration of stress at the domain wall, but a considerable level of focusing is preserved even for thicker hinges.

We now proceed to demonstrate the stress-focusing behavior experimentally. 
Using the same geometry shown in Fig.~\ref{IdealBC}, we cut a lattice from a 3.175 mm-thick sheet of stainless steel using water-jet cutting.
Fig.~\ref{CenteredExpModel}(a) shows the experimental setup for tensile tests in an MTS servo-hydraulic load frame.
It is important to recognize that, in the loading configuration considered in Fig.~\ref{IdealBC}, which is ideal for demonstrating the focusing effect, a distributed force is applied along the entire loaded edge, while operating with a load frame requires that the load is prescribed as a concentrated force. 
In order to mitigate this discrepancy, we introduce a modification of the specimen design, whereby we bound the lattice with two strips of solid material (geometrically stiffer in-plane than the lattice) at the top and bottom edges, to distribute the concentrated load evenly along the edges.
At the center of the top edge, the specimen is fixed to the load frame through a load cell, which monitors the axial force, and a gripper that is secured to the lattice through bolted connections. 
At the center of the bottom edge, the lattice is connected to an actuating piston through a second gripper. 
Four strain gages are attached to the hinge ligaments of four cells located at increasing distance from the domain wall, as shown in Fig.~\ref{CenteredExpModel}(b), to capture the decay of axial strain from the domain wall into the bulk. 
While this setup approximates the conditions of Fig.~\ref{IdealBC}, the point load is expected to slightly bias the loading process along the mid line despite the presence of the solid layer. 
To capture these conditions in the numerics, we update the FEM model as shown in Fig.~\ref{CenteredExpModel}(c), where we explicitly model the holes for the bolted connections and apply traction boundary conditions at these sites.
At the bottom edge we apply an axial load along the bottom halves of the hole perimeters to model the way in which the bolted connections transfer load from the loading piston to the lattice.
The material properties are those of the 304 stainless steel used for the specimen: Young's modulus E = 200 GPa and Poisson's ratio $\nu = 0.29$. Fig.~\ref{CenteredExpModel}(c) shows levels of stress focusing similar to Fig.~\ref{IdealBC}(a). 
The load is capped at 1700 N to avoid inducing yield in the lattice; as shown in Fig.~\ref{CenteredExpModel}(c), the maximum stress is ~200 MPa, not exceeding the approximate yield stress of 215 MPa for a typical 304 stainless steel.

\begin{figure*}[tb]
	\centering
	\includegraphics[width=1\textwidth]{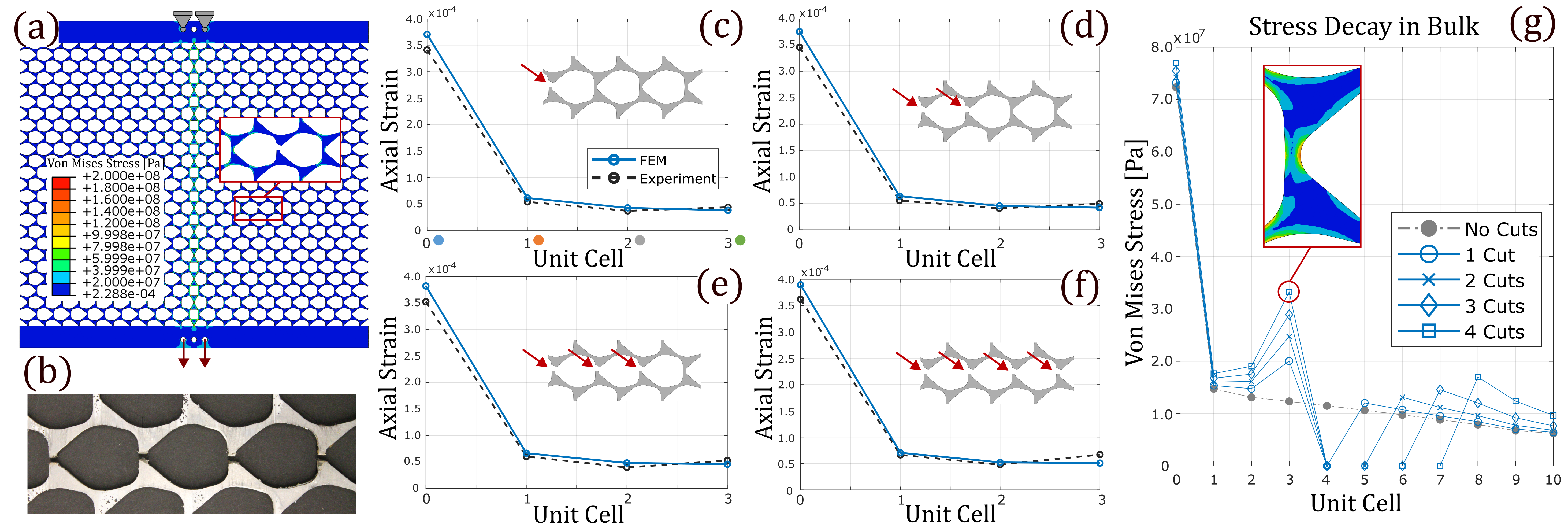}
	\caption{(a) Von Mises stress field of lattice with cut at a hinge. (b) Close-up of specimen with four cuts, mimicking conditions of an effective horizontal crack. (c-f) Decay of axial strain in the hinge ligaments from the domain wall into the bulk, for increasing number of cuts; comparison of FEM calculations and experimental strain gage readings. Numerics and experiments are in agreement and indicate robust focusing at the domain wall even in the presence of multiple cuts, and consequently protection of the damage zone. (g) Plot of von Mises stress vs. unit cell index for the lattice with and without cuts, showing larger sensitivity of von  Mises stress to the cuts, compared to axial strain.} \label{CutLattice}
\end{figure*}

We conduct the experiments by loading the lattice to 1700 N at a rate of 20 N/s, while sampling the relative change in voltage of the strain gages at 2 Hz. 
In post-processing, we convert the change in the voltage outputted by the strain gages to a measure of axial strain, $\epsilon_{yy}$, through shunt calibration and the following formula:
\begin{equation}
    \epsilon_{yy} = \frac{R_g}{V_i*GF*(R_c+R_g)}*\Delta V
\end{equation}
where the resistance of the strain gage $R_g = 120.3~\Omega$, the resistance of the shunt $R_c = 174800~\Omega$, the gage factor $GF = 2.09$, and the input voltage $V_i = 1.215~V$. 
$\Delta V$ is the change in voltage measured in the gage. 
Fig.~\ref{CenteredExpModel}(d) shows the history of axial strain in the gage located at the domain wall (cell 0) vs. the applied force, which grows linearly. 
The nearly perfect overlap of the loading and unloading curves suggests no hysteresis, confirming that the process remains in the elastic regime.
The imperfect connections between the specimen and the load frame, which are especially relevant at low load levels (when the bolt connections are loose due to their fabrication tolerances) inevitably result in some nonlinearity in the strain-force curves, shown in Fig.~\ref{CenteredExpModel}(e).
To filter out these spurious effects and probe the gages in the appropriate linear regime, we extract the slope of the final ~10\% of the curve and extrapolate it over the entire load history. 
To extract analogous information for axial strains in the hinges from the numerics, we proceed as illustrated in Fig.~\ref{CenteredExpModel}(c): for each hinge, we average strain values at the nodes of a portion of the mesh inside a 2 $\times$ 1 mm rectangle corresponding to the area that is covered by the strain gage in the specimen.
We compare experimental and numerical strains in Fig.~\ref{CenteredExpModel}(f) and we observe an overall excellent agreement, both in terms of decay rate, which is nearly perfect, and in terms of quantitative strain estimates, whose difference is less than 8\%.

\begin{figure*}[tb]
	\centering
	\includegraphics[width=1\textwidth]{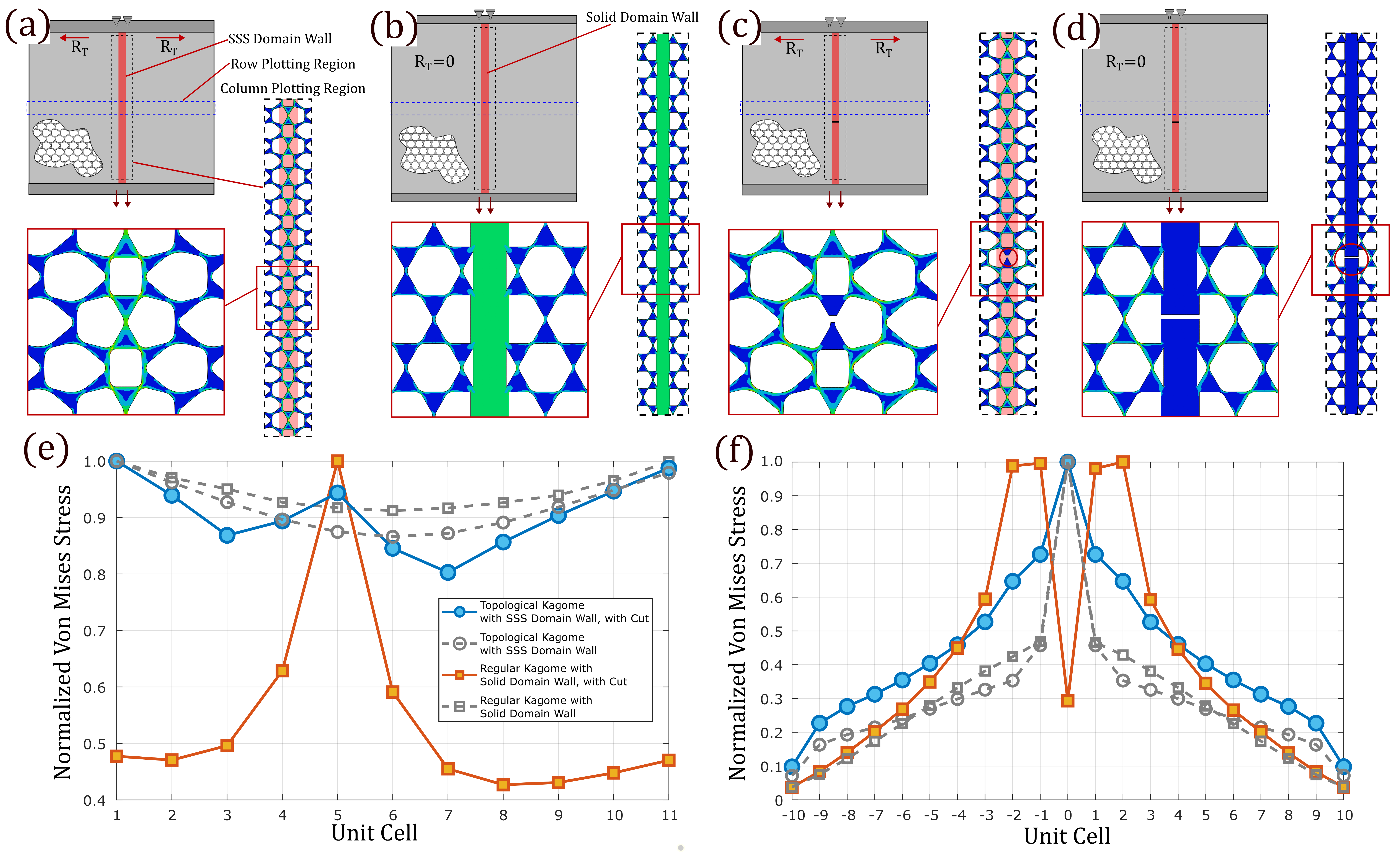}
	\caption{(a) Schematic of topological kagome lattice with SSS domain wall. Insets show the domain wall region and close-up of cells directly adjacent to the domain wall. Black and blue dashed boxes identify regions of stress sampling for plots in (e) and (f), respectively. (b) Schematic of unpolarized regular kagome lattice with a solid wall along its center axis. Insets show a concentration of stress along the domain wall. (c) Schematic of topological kagome lattice with SSS domain wall with a cut introduced at row 5. Insets show the exact location of the cut and the evolution of the stress field around it. The SSS domain wall keeps focusing stress despite the cut. (d) Schematic of the regular kagome lattice with solid wall and cut introduced at row 5. Insets show significant concentration of stress at unit cells directly adjacent to the cut. The domain wall does not focus stress effectively along its length. (e) Plot of (normalized) cell-averaged von Mises stress vs. unit cell index along vertical sampling region. (f) Plot of (normalized) cell-averaged von Mises stress vs. horizontal unit cell index.} \label{DomainWall}
\end{figure*}

We recognize that, in the lattice configuration considered so far, the applied load is coaxial to the domain wall.
This observation warrants the question whether this selection of boundary conditions exaggerates the results, making the observed strain localization along the center line a trivial artifact of the load axis selection.
To prove that the localization is non-trivial and actually resulting from the lattice polarization, we consider a modified configuration in which the load is applied with an eccentricity from the domain wall. 
Fig.~\ref{OffsetExperiment}(a) shows the updated FEM model using the same material properties and boundary conditions used before. 
Looking at the field of von Mises stress, we still observe a substantial localization at the domain wall. 
For a clearer comparison, in Fig.~\ref{OffsetExperiment}(b), for both loading configurations, we sample the axial displacement, von Mises stress, and axial strain, averaged over the cell, and plot it against the cell index along the row of unit cells enclosed in the dashed box.
Unlike the centered loading case, the offset loading produces a displacement field that grows linearly from left to right.
However, while globally the von Mises stress and axial strain fields also display similar trends, they both still feature a clear peak at the domain wall (unit cell 0). 
To reproduce the same loading conditions in the experiment, we leverage the solid edges at the top and bottom to introduce additional off-center holes to install the lattice in the offset configuration, as shown in the inset in Fig.~\ref{OffsetExperiment}(c).
The strain values measured by the gages are plotted in Fig.~\ref{OffsetExperiment}(c) alongside the values inferred from FEM.
Here again, numerics and experiments show good qualitative agreement and confirm the persistence of strong localization at the domain wall. 
For completeness, we report some appreciable quantitative discrepancy between experimental readings and numerics, incidentally mostly occurring at low levels of strain, where the sensitivity of the gages and other non-idealities in the measurements may more heavily affect the readings. 
However, these deviations do not make the emergence of the localization any less evident.

\section{Evidence of topological protection against damage}

As our next step, we want to verify experimentally the robustness of the stress focusing against the introduction of defects in the bulk.
To this end, we conduct a series of finite element analyses and experiments on configurations in which we introduce a variety of cuts in the hinge ligaments (akin to severed bonds in the ideal treatment in~\cite{Zhang_Fracture_2018}) to capture the onset of damage zones with progressively increasing size.
Starting with one cut directly adjacent to the fourth strain gage, we progressively add cuts to increase the length of an ``effective crack." 
The finite element model of the lattice with a single cut is shown in Fig.~\ref{CutLattice}(a), and a close-up detail of the specimen showing all four cut hinges is shown in Fig.~\ref{CutLattice}(b). 

\begin{figure*}[t]
	\centering
	\includegraphics[width=1\textwidth]{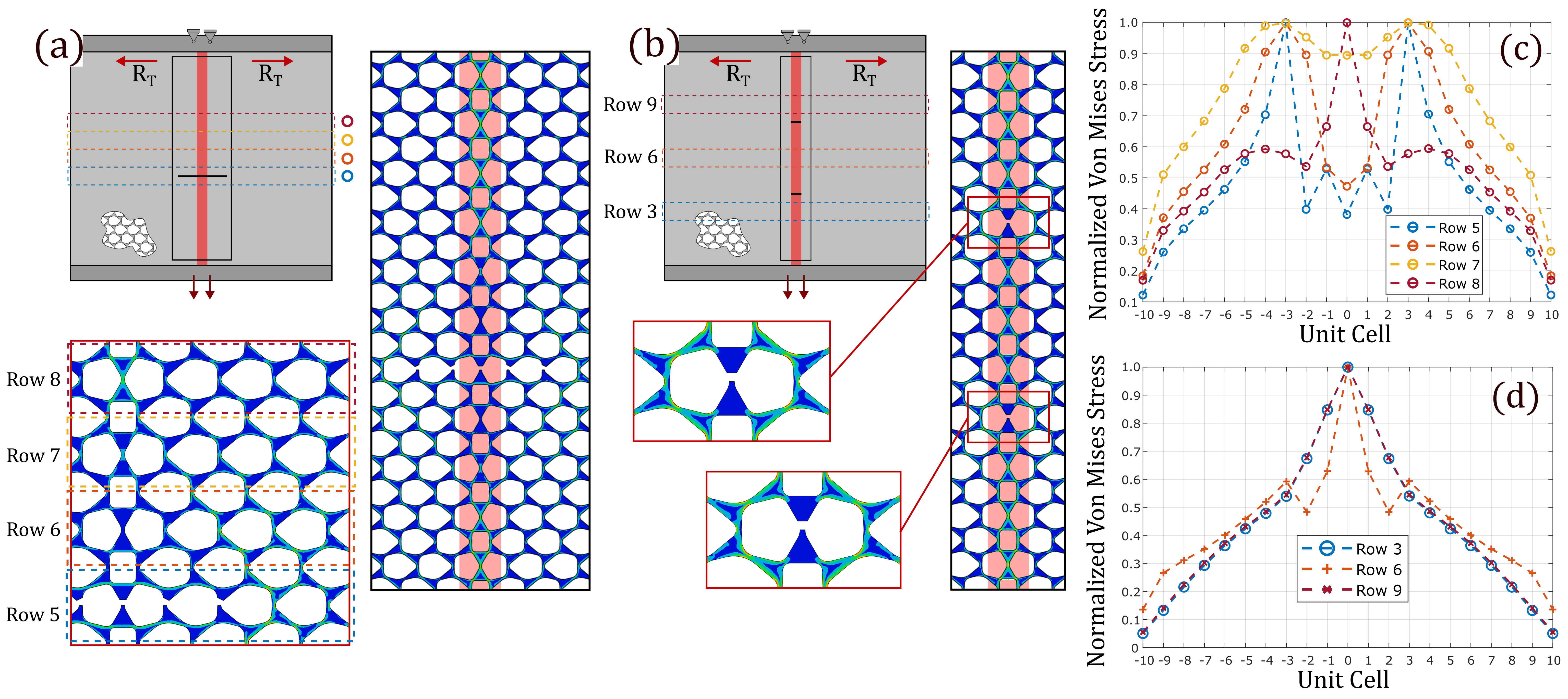}
	\caption{(a) Schematic and FEM model of the lattice with long cut across the SSS domain wall. Boxes mark rows sampled for the von Mises stress. Insets show evolution of stress field in a large diamond-like pattern around the cut. (b) Schematic and FEM model of lattice with two domain wall cuts highlighting persistent stress focusing along all the uncut portions of the domain wall, despite the loss of continuity. (c) Plot of domain wall-normalized von Mises stress vs. unit cell for the boxed rows, showing migration of highest stress away from the domain wall. At the highest sampled row, at the tip of the diamond region, the stress concentration returns to the domain wall. (d) Plot of stress sampled in the three boxed rows, confirming strong localization in each portion of the domain wall.} \label{LongCut_TwoCuts} 
\end{figure*}

From the von Mises stress field in Fig.~\ref{CutLattice}(a), we observe that significant stress focusing persists along the SSS domain wall despite the addition of the cut.
In Fig.~\ref{CutLattice}(c-f), we increase the effective length of the ``crack."
For each case we plot the axial strain in the hinge ligaments, as extracted from the strain gages and computed from the FEM calculations.
Numerics and experiments are in agreement and capture a strain landscape in which the focusing at the domain wall remains sharp and largely unaffected by the cuts.
This result demonstrates robustness of the focusing behavior against defects, an effect afforded by the topological protection inherent to the lattice.
The numerics allow us to take a look at other local mechanisms of deformation of the hinges beyond the axial strain that the  gages can capture.
In Fig.~\ref{CutLattice}(g), we plot the magnitude of von Mises stress in the hinges for the right half of the lattice.
For this metric, we do observe a more significant change near the crack tip as more cuts are introduced.
What this result suggests is that other components of the stress tensor aggregated in the von Mises stress, other than the axial one, start playing a major role in the deformation of the hinges. 
These other components of stress are likely caused by the bending mechanisms related to the opening of the cut under uniaxial tension of the lattice. 
However, it still appears that four cuts are not enough to completely overwhelm the stress focusing at the SSS domain wall and nullify the topological protection of the crack.

\section{Comparison with bar-like reinforcements}

The SSS domain wall studied thus far acts as a sort of ``reinforcement," whereby stress and strain are transferred away from critical locations where we want to avoid potential failure (e.g., the bulk of the lattice) towards a pre-designed location. 
This observation begs for the following question: what aspects, and how much of the strength of the observed effect, can be attributed to the topological polarization, rather than to a generic capability to focus stress displayed by a broader class of trivial mechanical systems endowed with some form of reinforcement?
In other words, how does an SSS domain wall compare against a more conventional, non-polarized lattice featuring a solid bar-like or fiber-like reinforcement? 
To address this question, we consider a regular kagome lattice, which is non-polarized, and we endow it with a vertical \textit{solid wall} in the form of a slender bar, as shown in Fig.~\ref{DomainWall}(b); this effectively partitions the lattice into two subdomains, to mimic the conditions considered so far for the topological case. 
While this wall is not topological, i.e., it does not bound regions with different topological polarizations, its fiber-like morphology and orientation with respect to the direction of loading arguably make it an effective pathway for localizing stress.
As shown in the insets in Fig.~\ref{DomainWall}(b), the stress indeed localizes along the length of the solid wall.
If we compare this outcome against the polarized kagome lattice studied so far, reported side-by-side in Fig.~\ref{DomainWall}(a), one would be tempted to conclude that the lattice polarization does not offer major advantages. An immediate counterargument could be that, in Fig.~\ref{DomainWall}(a), we preserve a Maxwell cellular architecture over the whole structure, albeit with a local change in cell shape at the domain wall; this significantly limits any detrimental impact introduced by the domain wall on the mechanical properties (e.g., effective elastic moduli) and elastodynamic behavior (e.g., domain wall scattering) of the structure, which are instead severe with a solid interface. 

There are other considerations, more germane to the discussion on topological polarization, that provide additional arguments in support of the superiority of the topological domain wall. 
To elucidate these, we proceed in a similar manner to the previous section and we introduce cuts along the walls for both cases, as shown in Fig.~\ref{DomainWall}(c) and (d).
For the SSS domain wall (Fig.~\ref{DomainWall}(c)), like the previous results, we observe a local reconfiguration of the stress in the neighborhood of the cut, where the stress flow necessarily extends to the cells bounding the domain wall and bypasses the cut. 
However, the entire domain wall continues to act as an effective stress guide, despite the cut. 
In contrast, the same cut applied to the solid wall makes the stress localized to the interface drop drastically, as shown in Fig.~\ref{DomainWall}(d). 
This is a powerful demonstration of topological protection: as long as the polarization of the bulk is preserved, the domain wall conserves its attributes despite major local modifications of its geometry. 
The fact that focusing is maintained even in these drastic cases, with the stress localization systematically bounding the domain wall despite its snaky profile, represents another eloquent demonstration of the topological protection.
In order to better quantify these different effects, in Fig.~\ref{DomainWall}(e) we calculate the average von Mises stress in one-cell-tall sections inside the vertical region (marked by a black dashed line) and we plot it (normalized by the highest value) against the vertical unit cell index. 
This metric is supposed to capture the degree of homogeneity of the stress along the domain wall. 
Clearly, without the cut, the average stress is fairly constant with either domain wall (gray curves). However, when we introduce the cuts, the SSS domain wall continues to focus stress (blue circular markers), while for the solid wall, all the stress peaks near the cut (orange square markers). 
Another difference concerns the strength of the decay into the bulk featured by the two domain walls, which is captured by the graph in Fig.~\ref{DomainWall}(f), in which we plot the average cell stress against the cell index sampled along the horizontal region marked by the blue dashed line.
Clearly, for the SSS domain wall lattice, even in the presence of a cut, the stress remains effectively localized at the domain wall with sharp decay into the bulk.
In contrast, while the regular kagome lattice shows good stress focusing when uncut, the cut lattice actually shows a significant decrease in performance; the maximum stress is no longer located at the domain wall, rather it is concentrated in the unit cells immediately adjacent to the domain wall.

\section{Further evidence of domain wall's topological attributes}

We further bolster the evidence of the topological protection granted by the SSS domain wall with two more FEM models: one featuring a longer cut extending from the domain wall into the bulk, and the other featuring multiple cuts along it. 
Starting from the lattice in Fig.~\ref{DomainWall}(c), we extend the cut through an additional two unit cells on both sides of the domain wall, as shown in Fig.~\ref{LongCut_TwoCuts}(a). 
As visible on the von Mises stress field map, the introduction of this cut dramatically alters the morphology of the stress field about the domain wall to that of a large diamond-like shape. 
To quantitatively document the shape of the localization profile, we plot the stress (averaged and normalized as in previous cases) along four rows of cells at different distances from the cut marked by the dashed boxes in Fig.~\ref{LongCut_TwoCuts}(a). 
The plot in Fig.~\ref{LongCut_TwoCuts}(c) shows a consistent pattern, whereby the stress peaks progressively migrate outwards as we move closer to the cut rows, with a drop in stress over a larger set of cells across the cut region. 
The fact that the localization always occurs at the domain wall (or at the closest solid feature), regardless of the shape of the interface, confirms that such localization is guaranteed regardless of the domain wall morphology -  a testament to the power of topological protection. 
For the second model we introduce two short cuts to the lattice at separate spots along the domain wall as shown in Fig.~\ref{LongCut_TwoCuts}(b). 
We observe that, while the stress field detours around both cuts, the localization remains pronounced along each uncut portion of the domain wall, even \textit{between} cuts, despite the interruption of the effective continuity of the interface. 
In essence, each surviving portion of the domain wall still acts as an effective pathway for the localization of stress. 
In Fig.~\ref{LongCut_TwoCuts}(d), we see that the normalized stress plotted along the rows marked by the shaded boxes undergoes the same evolution, peaking at the domain wall and rapidly decaying into the bulk.

\section{Conclusion}

We have demonstrated topologically protected stress-focusing behavior in a topological metamaterial with finite-thickness hinges, through both simulations and experiments.
The selected lattice configuration is found to feature significant stress focusing along the domain wall between the subdomains.
Experiments with strain gages capture a sharp decay of axial strain measured in the hinges from the domain wall into the bulk. 
This strain-focusing is observed even under the action of highly eccentric loading and, more importantly, in the presence of defects in the bulk. 
The experiments demonstrate that the SSS domain wall de facto protects the lattice from the potential of damage and fracture by mitigating the stress localization in the neighborhood of stress hot spots such as the tips of cracks. 
Finally, through numerics, we also have been able to demonstrate the superiority of the topological kagome paradigm over a more generic type of fiber-reinforced structure.

\section{Acknowledgments}

The authors acknowledge the support of the National Science Foundation (award CMMI-2027000).  The authors are grateful to the Minnesota Supercomputing Institute for access to software and computational resources used in the simulations. The authors are also grateful to Pouyan Asem for his help with the experimental setup, and to Harry Liu for assistance with the theoretical background. S.G. is especially grateful to Kuan Zhang for his deep insight and discussions on the problem. 

\bibliography{example.bib} 

\end{document}